# Replacing Network Coordinate System with Internet Delay Matrix Service (IDMS): A Case Study in Chinese Internet

Gang Wang[*], Chunhong Zhang, Xiaofeng Qiu, Zhimin Zeng

Mobile Life and New Media Lab, Beijing University of Posts and Telecommunications, 100876, Beijing, China .

## Abstract

Network distance (Round Trip Time, RTT) is an important parameter for many Internet distributed systems to optimize their performances. Network Coordinate System (NCS) is assumed as a lightweight and scalable mechanism to predict network distance between any two Internet hosts without explicit measurements. Though many NCSes have been proposed in the literatures, they are not satisfactory in terms of accuracy. In this paper, we propose to use delay matrix to replace the NCS. This paper makes three contributions. First, we show that not all the hosts need an independent network coordinate (NC). On the contrary, most hosts can be represented as one or several nodes in NCS. Second, we present an Internet Delay Matrix Service (IDMS) for representing network distances without explicit measurements in Internet. Third, we describe two delay matrices—up-to-date delay matrix (UDM) and previous delay matrices (PDM)—for representing the network distances. Extensive simulations on our collected Chinese Internet data sets show that IDMS is an accurate, efficient, scalable and practical method for Chinese Internet. The performance of IDMS is better than other existing approaches.

## 1. Introduction

Network distance is an important parameter for optimizing Internet distributed systems such as content distribution networks (CDN), multiplayer online games [1], peer-to-peer (P2P) distributed hash tables (DHT) [2-3], P2P voice-over-IP (VoIP) relay node selection [4], application layer multicast [5], file-sharing based P2P [6]. Network distance can be measured directly via end-to-end probes such as *PING*, inferred indirectly such as *KING* [7], or estimated using NCS. To obtain accurate knowledge of network distances from active network measurements is expensive and time-consuming, especially when the size of network is very large. Thus, *PING* and *KING* are not scalable mechanisms.

Thus, designing a model to predict unknown network distances from limited samples of Internet measurements has become a hot research topic in the research community in recent years. Many NCSes have been proposed in the literatures. Most NCSes are based on Euclidean distance model or matrix factorization model.

For the Euclidean distance model such as GNP [8], Vivaldi [9], ICS [10], VL [11], NPS [12], PIC [13], each host is embedded in a low dimensional space called network coordinate and the Euclidean distance between the coordinates of two hosts is used to estimate the RTT between the two hosts. Unfortunately, Euclidean distance model has two common drawbacks. One is that it is incapable to represent triangle inequality violation (TIV) in their metric spaces which is a common phenomenon on the Internet. The other is that it cannot represent networks with complex routing policies such as asymmetric routing [20] since Euclidean distances are inherently symmetric. Therefore, NCSes based Euclidean distance model can not achieve high prediction accuracy.

For the matrix factorization model such as IDES [22] and Phoenix [23], each host has two *d*-dimensional row vectors. One is the incoming vector. The other is outgoing vector. The predicted distance from host *i* to host *j* is the dot product between the outgoing vector of host *i* and the incoming vector of host *j*. The advantage of matrix factorization model is that it can overcome certain limitations of Euclidean distance model. For example, there is no constraint to satisfy the triangle inequality. The distance computed model of matrix factorization model can represent asymmetric routing. However, one drawback of some NCSes based matrix factorization model will give negative predicted distance which is harmful to many classical algorithms such as Dijkstra algorithm since they are based on the assumption of non-negative distances. The other drawback is the propagation of the prediction error can impact prediction accuracy [23]. Though there are some approaches to improve the prediction accuracy of these NCSes, the prediction accuracy and the TIV problem are not fundamentally solved.

[*] Corresponding author. Tel.: +86 010 62253599; fax: +86 010 62253599.
E-mail address: wifnstone@gmail.com (Gang Wang), zhangch@bupt.edu.cn (Chunhong Zhang), qiuxiaofeng@gmail.com(Xiaofeng Qiu), zengzm@bupt.edu.cn (Zhimin Zeng).

For most NCSes, they treat Internet as a black box and do not exploit the properties of Internet architecture. The main role of NCSes is to provide location service, but according to a measurement study of a popular BT system Azureus [26], Vivaldi is not suitable for selecting close-by nodes which is very important to reduce cross-ISP traffic in P2P systems. Since most NCSes can not perfectly handle the TIV problem, exploiting NCSes to find detour path [27] for delay-sensitive applications may be very difficulty.

The motivation of this paper is to explore a new method to overcome the limitations of NCSes and answer the following questions.
1. Does each host need an independent network coordinate (NC)?
2. How to design a distributed and efficient location service without explicit measurements in Internet to remedy the limitations of NCSes?

To answer these questions, we conducted intensive Internet measurements to study the delay relationship among hosts in Chinese Internet. Through measurement study, we demonstrate that the path delays within the same Internet Service Provider (ISP) or Autonomous Systems (AS) are very small and vary slightly. TIV phenomenon will rarely occur within the same ISP or AS. Thus, hosts located in the same ISP or AS may be represented as one node in the NCS. The most congested links in terms of delay are the links between some large ISPs. Therefore, NCSes should pay more attention to the delays among ISPs or ASes since the practical detour paths may cross different ISPs or ASes. The path delay will vary in different time period and show diurnal behavior which has peak value in the work hour and evening and low value in the middle night to morning. The diurnal behavior means that the delay matrices of Internet are similar in the same time period of every day. Thus, the previous delay matrices can be used to estimate the future delay matrices with the same time period.

Exploring these findings we bring forward a previously unexplored approach and design Internet Delay Matrix Service (IDMS) to representing network distances without explicit measurements in Internet. The main idea is that IDMS will use a small delay matrix to represent the network distance. IDMS will periodically measure and generate a small delay matrix called up-to-date delay matrix (UDM) in different time periods. This delay matrix will be periodically distributed to some powerful nodes in the network. IDMS also introduces previous delay matrix (PDM) which was measured in previous days. The PDMs can be stored in every host since PDMs will not be updated frequently. IDMS directly use the delay matrix to represent the network distance, so the TIV problem can be perfectly solved. In IDMS, there is only one delay matrix in one specific time period, so the node who has stored the delay matrix can obtain the knowledge of network distance of the whole network. This property can benefit for detour path selection. Extensive simulations on our collected Chinese Internet data sets show that IDMS is an accurate, efficient, scalable and practical method for Chinese Internet. The performance of IDMS is better than other existing approaches.

The rest of this paper is organized as follows. In section 2, some existing NCSes based on Euclidean distance model and matrix factorization model are reviewed. Section 3 analyzes the properties of delay space of Chinese Internet and its usability in IDMS. Our proposed IDMS is provided in section 4. We intensively evaluate IDMS and compare it with Phoenix by trace-driven simulation in section 5. We summarize the paper in section 6.

## 2. Related Work

### 2.1 Active Measurement

To obtain the knowledge of network distances, one method is to use active measurement where each node should active measure all the other nodes to obtain the network distances. For active measurement nodes can use *PING* to directly probe other nodes, or use *KING* to infer indirectly. *PING* only can obtain the network distances from one node to all the other nodes. For one node, using *PING* can not generate the whole network distances. *KING* is a very useful technology that can run on one node to estimate the network distance of any pair of nodes. However, active measurement is expensive and time-consuming, especially when the size of network is very large. For example, for a network with $N$ hosts, each host requires $O(N)$ measurements to obtain the network distances between itself and all the other hosts, while the whole network will generate $O(N^2)$ measurements. For one host, if it requires to obtain the whole network distance it will require $O(N^2)$ measurements. Such measurements would not only impose heavy load on all the hosts but also consume huge network bandwidth. Thus, *PING* and *KING* are not scalable mechanisms when the size of network is very large.

## 2.2 Network Coordinate System based on Euclidean Distance Model

Since active measurement can not scale to large-scale network, how to design a model to predict unknown network distances from limited samples of Internet measurements has become a hot research topic in the research community in recent years. Many NCSes have been proposed in the literatures. Most NCSes are based on Euclidean distance model or matrix factorization model.

For the Euclidean distance model, each host is embedded in a low dimensional space called network coordinate to represent its position in the Euclidean space, and the Euclidean distance between the coordinates of two hosts is used to estimate the RTT between the two hosts. Unfortunately, Euclidean distance model has two common drawbacks. One is that it is incapable to represent Triangle Inequality Violations (TIVs) in their metric spaces. Many studies [14], [15], [16], [17], [18] have reported that TIV is a common phenomenon on the Internet. Triangle Inequality Violation (TIV) is the phenomenon that the direct path is longer than the relay path through an intermediary in terms of delay. For example, nodes *A, B, C* form an invalid triangle if $L(A,B) > L(A,C) + L(C,B)$, where $L(X,Y)$ is delay between node $X$ and node $Y$. The other is that it cannot represent networks with complex routing policies such as asymmetric routing [20] since Euclidean distances are inherently symmetric. Therefore, NCSes based Euclidean distance model can not achieve high prediction accuracy.

To overcome the problem of TIV, the authors [16] proposed to avoid coordinates updated from serious TIV edges. For example, if one edge shows serious TIV, then this edge will not be used to compute the coordinates. The authors [19] proposed to partition the TIV links into different autonomous NCS in order to make as many as TIVs inherently embeddable in metric space. Another interesting paper called PeerWise [24] is not to improve the accuracy of NCS but to explore the high embedding errors of coordinates to search detours. However, PeerWise introduced extra overhead and may not find the suitable detours. For example, for a P2P-VoIP application, if the delay of detour path is not small than 150ms, the detour path will not satisfy VoIP quality requirement. However, these methods to overcome TIV didn't solve the TIV problem fundamentally.

Currently, Vivaldi has been widely deployed and implemented as a basic building block in most Internet systems, such as Peerwise Overlay [24], Census [25], and Azureus BitTorrent [26], but according to a measurement study of Azureus [26] Vivaldi is not suitable for selecting close-by nodes (within same country or ISP). For example, the authors [26] plotted the network coordinates version (Figure 5) of all countries. However, it is not possible to distinguish between countries or even continents.

## 2.2 Network Coordinate System based on Matrix Factorization Model

The IDES based matrix factorization model was first introduced in paper [22] to represent and estimate distances in large-scale networks. The matrix factorization model is base on the observation that if two nearby hosts have similar network distances to all the other hosts in the network, then their corresponding rows in the distance matrix will be nearly identical. The essential idea matrix factorization model is that if there are many equal or nearly equal rows in a large distance matrix, the distance matrix can be approximately factorized into two smaller matrices by methods such as Singular Value Decomposition (SVD) or Non-negative Matrix Factorization (NMF) [21]. Different from Euclidean distance model in IDES each host has two *d*-dimensional row vectors. One is the incoming vector. The other is outgoing vector. The predicted distance from host *i* to host *j* is the dot product between the outgoing vector of host *i* and the incoming vector of host *j*.

The advantage of matrix factorization model is that it can overcome certain limitations of Euclidean distance model. For example, distances computed in IDES are not constrained to satisfy triangle inequality. The distance computed model of IDES can represent asymmetric routing. One drawback of matrix factorization model is that IDES base SVD will give negative predicted distance which is harmful to many classical algorithms such as Dijkstra algorithm since they are based on the assumption of non-negative distances. The other drawback is the propagation of the prediction error can impact prediction accuracy [23]. According to [23], IDES can not achieve higher prediction accuracy than Euclidean distance model.

Recently, another matrix factorization model Phoenix [23] was proposed to overcome the drawbacks of IDES. Phoenix introduced a weight-based mechanism for the NC calculation which can improve the predict accuracy. The main idea is that in the NC calculation accurate reference NCs are given higher weight and inaccurate reference NCs are given lower weight or abandon. Phoenix can provide more accurate prediction accuracy and handle TIV better than other NCSes, but Phoenix is still not an ultimate solution since the prediction accuracy and the TIV problem are not fundamentally solved.

## 2.3 Absolute Error and Relative Error of NCS

The prediction accuracy of a NCS is often evaluated by absolute error (AE) and relative error (RE) [9], [15], [12], [22]. Suppose there are $N$ hosts in the network. Let $L(i,j)$ be the measured RTT between host $i$ and host $j$, and $L^E(i,j)$ be the predicted distance between them.

Absolute error (AE) of one link is defined as:

$$AE(i,j) = |L(i,j) - L^E(i,j)| \tag{1}$$

Relative Error (RE) of one link is defined as:

$$RE(i,j) = \frac{|L(i,j) - L^E(i,j)|}{L(i,j)} \tag{2}$$

For the whole network, most NCSes seek to minimize the sum of AE of all links. For one link, smaller AE or RE indicates higher prediction accuracy.

## 3. Properties of Delay Space of Chinese Internet and its Usability in IDMS

Knowing the properties of delay space of Internet is very important to design a high performance of Internet location service. Though most public data sets of delay space of Internet such as King data set (1740 nodes) [28] are available on the Internet, one problem of these data sets is that they can not show the dynamic properties of delay space of Internet. For most public data sets it only contains one delay matrix where the final RTT between two hosts is the median or minimum value of RTTs which was measured over long periods of time such as days or even weeks. Therefore, these data sets are not suitable for analyzing the dynamic properties of delay space of Internet.

Exploiting the properties of delay space of Internet is our focus to design IDMS in this paper. In this section we will analyze the properties of delay space of Chinese Internet. We need some continuous delay matrices which are measured with small time granularity. For example, if the time granularity is one hour to generate one delay matrix of Chinese Internet, the continuous delay matrices will contain 24 delay matrices in one day. We also need to analyze the delay relationship among ASes in Chinese Internet.

The measurement methodology has been reported in our early paper [18, 29]. In the measurement, we use the *KING* tool to measure the delay space of Chinese Internet. We set the sampling interval of measurement as 1-hour. We run *KING* from a computer at our university. The median of delay sample is used for each pair of DNSes in each sampling interval. The whole measurement lasts 14 days (day 1-day 14). In each day the *KING* can obtain 24 delay matrices (329x329). Some properties of delay space of Chinese Internet have been reported in our early paper [18, 29]. Here, we only give the results and its usability in IDMS.

*Property 1: There are numerous TIVs in Chinese Internet.*

This property has been reported in our early paper [18]. To reserve as many TIVs as possible, IDMS adopts the delay matrix to represent the delay space of Chinese Internet.

*Property 2: The delay matrices with the same time period are highly similar in every day. The previous delay matrices can be used to estimate the future delay matrices.*

This property is similar to daily patterns. In order to analyze this property, we select two delay matrices measured in different time period from day 1 as the benchmark. The first delay matrix was measured between 5:00 and 6:00 which represents the least congested network scenario and is called LCN in this paper. The second delay matrix was measured between 21:00 and 22:00 which represents the most congested network scenario and is called MCN in this paper. We compare them with the other delay matrices (day 2-day 6, and day 14) with the same time period. Our comparison is based on AE and RE which are defined in formula 1 and formula 2.

Figure 1 plots the absolute errors in LCN. We observe that the delay matrices are highly similar. For about 95% of absolute errors are less than 20ms. It also shows that the absolute errors become big if the time span becomes big. However, it can also maintain high similarity. Figure 2 plots the relative errors in LCN. We can observe that over 60% of paths the relative errors are close to zero. Approximately 90% of relative errors are less than 0.3. These errors may not be very serious. While some errors are extremely large, we consider them as measurement problem. Since KING relies on two DNSes to measure the delay between them, the transient overload at the DNS servers is inevitable which can cause query delay as discussed in [17]. The results of MCN also present the similar results except that the AE and RE are slightly bigger. We also analyze the other delay matrices. The results are similar.

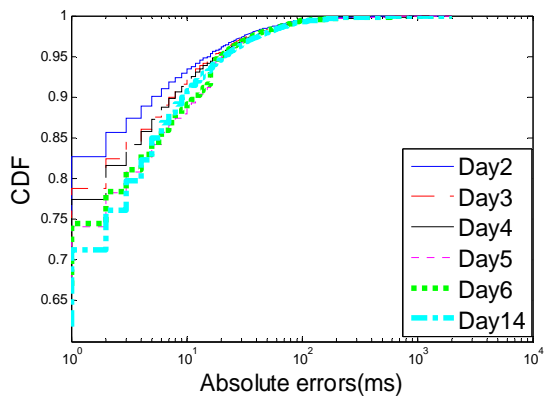
Fig. 1. Absolute errors of LCN

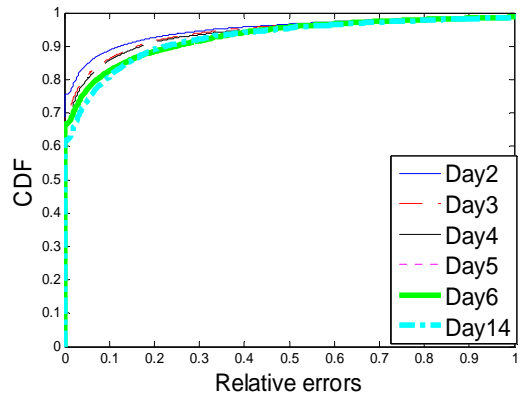
Fig. 2. Relative errors of LCN

The above observations suggest that the delay matrices with the same time period are highly similar. It means that the previous delay matrices can be used to estimate the future delay matrices with the same time period. In IDMS, it will use two types of delay matrix to represent the delay space of Chinese Internet. One is the up-to-data delay matrix (UDM) which is measured every hour. The other is the previous delay matrix (PDM) which was measured in previous days. PDM can also provide high accuracy since the delay matrices with the same time period are highly similar.

*Property 3: The path delay will vary in different time period. Some paths undergo large variation. Some paths undergo slight variation or no change.*

This property has been reported in our early paper [29]. This property is easy to understand. Some early studies have showed that the Internet traffic presents daily patterns, so the path latency may become high when the traffic increases. However, the increasing trend may not be the same. For example, when the links and the routers are not overloaded, the path latency will not increase apparently even if the traffic increases. The paths which undergo larger latency variation are mainly because some links or routers on the path are congested when there are more users using the Internet. Though the traffics become larger on almost all paths when more users use the Internet, the links or routers on some paths may not be congested. Therefore, the latencies of these paths vary slightly or do not change. If there are not larger queuing delays in the routers, then the path delays will be very small.

The above observations suggest that one delay matrix is not enough to represent the delay space since the path delay will vary in different time periods. IDMS will generate more delay matrices to represent the delay space of Chinese Internet.

*Property 4: The path delays within the same ISP vary slightly. The most congested links in terms of delay are the links between some large ISPs.*

This property has been reported in our early paper [29]. The path delays within the same ISP vary slightly and usually are less than 100ms or 150ms. The most congested links are the links among some large ISPs. For example, even if the source and destination are located in the same city, the path latency between China Telecom and China Unicom is also very large. We believe that this phenomenon is not the technology problem, but rather reflects complex commercial contractual relationship of these large ISPs, such as fierce competition.

The above observations suggest that not all the hosts need an independent network coordinate (NC). For example, the hosts in the same ISP are close to each other, so they only need to identify whether the hosts are in the same ISP or not. Since the hosts in the same ISP are close to each other, TIVs may seldom occur within the same ISP. For one ISP or AS, maybe one node or several nodes are enough to describe the relationship with other ISPs or ASes since their paths show the similar property (congested or not congested). In IDMS, the delay matrix only contains the path delays among ASes and abandons the path delays within the same AS. Thus, the size of the delay matrix of all hosts can be decreased. The size of delay matrix of IDMS is similar to the number of ASes in Chinese Internet. Therefore, the delay matrix can be easily measured and disseminated. The storage and process time required is also very small.

## 4. Design of IDMS in Chinese Internet

In the previous section, we have analyzed the properties of delay space of Chinese Internet. Motivated by the measurement studies, in this section we present the design of IDMS. Different from NCSes, in IDMS each host does not have an independent coordinate. On the contrary, IDMS will use a very small delay matrix to represent the delay matrix of all hosts in the system. Specifically, the delay matrix is composed of network distances among the ASes in Chinese Internet. For the hosts in the same AS, IDMS assign them one unique host identity (ID) prefix to distinguish the hosts in the other ASes. In IDMS, each host will store the same delay matrix. Each host will select one point in the delay matrix as its reference point. Therefore, the network distance between arbitrary two hosts can be directly looked up in the delay matrix.

### 4.1 System Model

We consider a network in China Internet consisting of a large collection of hosts. Based on the measurement studies, hosts in the same AS are close to each other. The path latencies within the same AS are very small and vary slightly. The path latencies among ASes can be estimated by direct IP routing between a pair of nodes in their corresponding ASes. With publicly available BGP tables and updates, such as RouteViews (http://www.routeviews.org/), it is easy to obtain an IP prefix to AS number (ASN) mapping table of Chinese Internet. And the IP prefix to ASN mapping table will not change frequently.

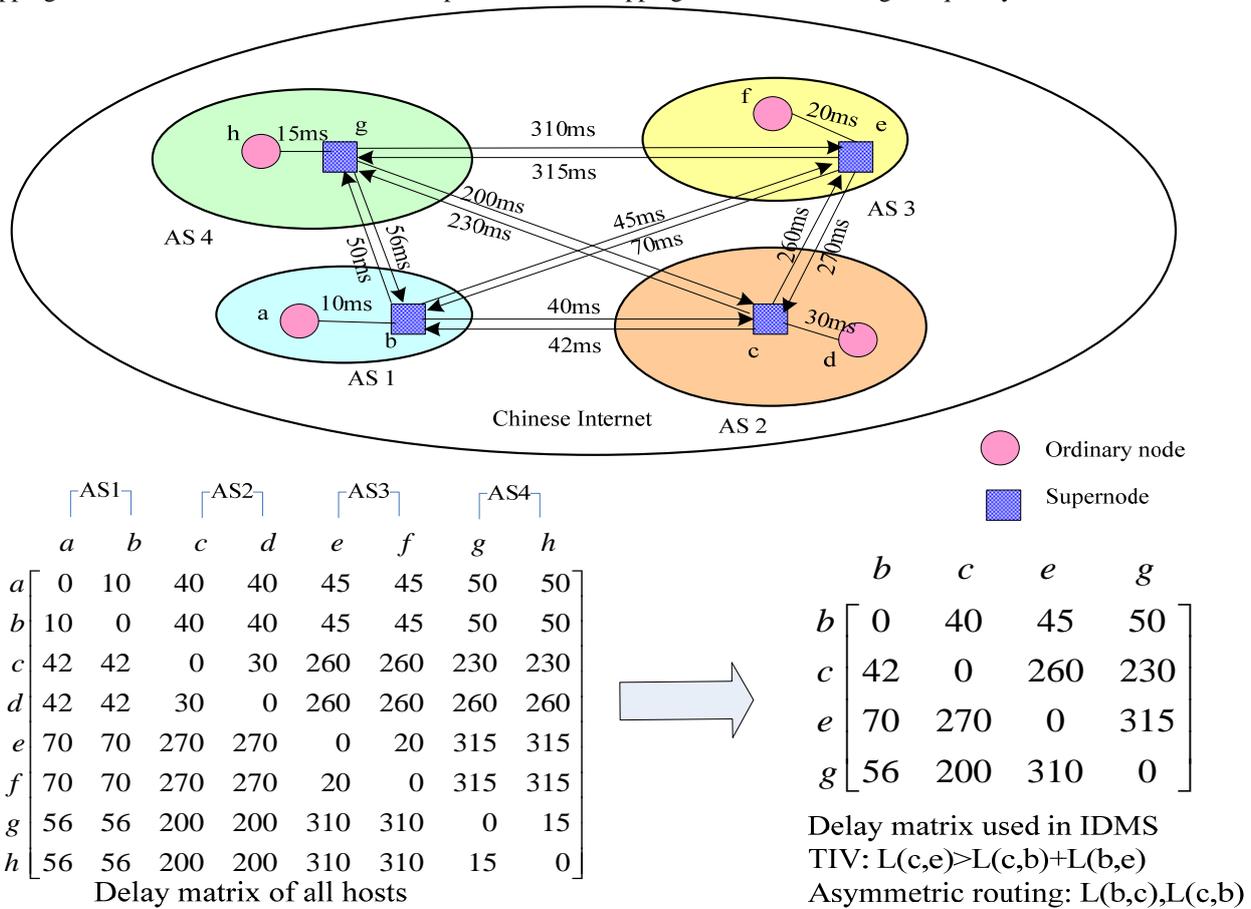

Fig. 3. **System model**

Figure 3 shows the system model used in IDMS, in which hosts are clustered based on ASN. In each ASN, a part of powerful hosts are selected as supernodes (SNs). The other hosts are ordinary nodes (ONs). The basic idea of IDMS is to use a very small delay matrix to replace the whole delay matrix of all hosts. For example, for a system with $N$ Internet hosts, the $N \times N$ Internet distance matrix of all hosts can be represented with a smaller $n \times n$ matrix ($n<<N$). Specifically, the delay matrix used in IDMS will focus on the delay relationship among ASes and only contains delays among ASes in Chinese Internet. The delays among hosts within one AS are abandoned since our measurement study shows that these delays are very small which can satisfy almost all Internet applications. In other word, all most all paths within one AS may not need to use detour paths to improve their performance. Therefore, in IDMS each host will store the same delay matrix. Since IDMS does not further process the delay matrix, the problems of TIV and asymmetric routing can be perfectly solved.

In IDMS, it defines two types of delay matrices. One is the update delay matrix (UDM) which is periodically measured and disseminated to all SNs. The other is the previous delay matrix (PDM) which can be generated by using the delay matrices measured in previous days. For any latency of PDM, the median value is used which is computed from several previous delay matrices measured in the same time. For example, if one path latencies were 51ms, 55ms and 63ms separately in three previous delay matrices with the same time period, then 55ms was used in PDM. UDM will be stored in each ON. In current design, IDMS will generate one UDM per hour. So in each day, IDMS will generate 24 UDMs. Each UDM and PDM is associated with one time stamp to distinguish.

## 4.2 System Architecture Overview

As shown in figure 4, the system architecture of IDMS is a two-tier architecture based on P2P, in which hosts are clustered based on ASN. Two-tier architecture has been demonstrated as one high-performance architecture for P2P system [31, 32]. It takes advantage of heterogeneity of capabilities across hosts and allows weak hosts and hosts behind NATs/firewalls to utilize the P2P services. Hosts in top tier are SNs, while hosts in bottom tier are ONs. To easy presentation we use the structured P2P network as an example which has been implemented in many P2P applications such as BitTorrent-based file sharing. SNs form one-hop Distributed Hash Table (called SN-DHT) such as D1HT [30] and ONs form multi-hop DHT (called ON-DHT) such as Chord [2]. Each ON is associated to at least two SNs to enhance the reliability.

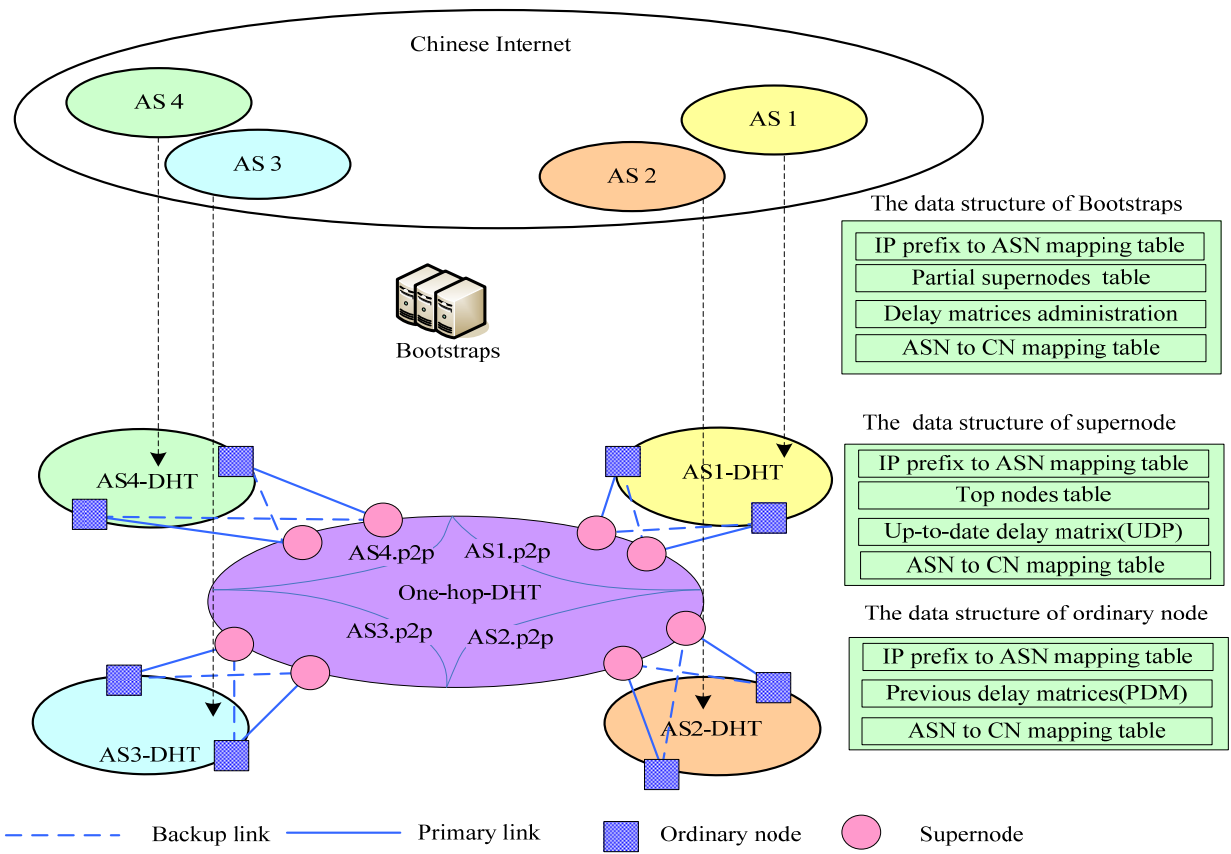

Fig. 4. **System architecture of IDMS**

### 4.2.1 DHT in IDMS

DHT can provide a service similar to the traditional hash table. In DHT overlay, each node has a unique numeric identifier (node ID) and each resource has a numeric identifier (key). Usually node IDs are mapped from the IP addresses and keys are mapped from the resource names with consistent hash function. Every node in DHT overlay maintains a partition of key space, and collectively all nodes maintain the whole hash table. Every node also maintains a routing table. These nodes collectively maintain and operate data based on DHT algorithms. With the routing table and routing algorithm, any node can insert/remove/retrieve (key, data) pairs in DHT. Compared with multi-hop DHT, every node in one-hop DHT maintains *N-1* routing table entries where *N* is the number of node in the system, while the typical routing table entries maintained by a node in multi-hop DHT is *O(logN)*. The maximum hop count of routing in one-hop DHT is one, while the average hop count in multi-hop DHT is *O(logN)*. Though the overhead of one-hop DHT is larger than that of multi-hop DHT, when the SNs are more reliable and the number of SNs is not very large one-hop DHT is more suitable for SNs network.

### 4.2.2 Locality-based cluster of IDMS

Approaches to utilize locality/topology information in DHT can reduce the latency per DHT hop and the DHT routing hop count, such as proximity routing [33] and topological/geographic layout [34-35]. Proximity routing selects the node with the minimum RTT as next hop from a set of neighbors closer to the destination. Topological/geographic layout exploits topological/geographic locality information into node ID generation.

Our previous work [35] proposed a locality-aware P2PSIP networks. The locality information used in that paper is based on country. This cluster method can be useful for the networks which cover many countries, but can not be used within one country. And our measurement study has showed that in Chinese Internet the hosts are not close to each other. IDMS should further cluster the hosts within one country.

For IDMS, there are two locality mechanisms. One is based on the country. The other is based on the ASN. As shown in figure 18, hosts from the same ASN are clustered together to construct a locality-aware DHT overlay. ONs are associated with SNs from the same ASN.

In IDMS, a locality identity (locality ID) is arranged as the first bits of the node ID. Its function is to construct the locality-aware DHT. Table 1 gives an example of constructing node ID. The locality ID is generated from country and ASN. Each country is mapped to a country number (CN) with a fixed mapping table. ASN is directly used in the locality ID. Thus, the CN and the ASN constitute the locality ID. The other part of node ID is mapped from the IP address with hash function.

**Table 1** Node ID

|  | Map(country) | ASN(4837) | Hash(IP address) |
|---|---|---|---|
| Node ID: | 01 | 1001011100101 | 10110110------ |

### 4.2.3 Definition of hosts in IDMS

In IDMS, like the well-known P2P-VoIP system, Skype, three types of hosts are defined as listed below:

**Bootstraps:** Bootstraps are the powerful, dedicated, and always-on servers. In IDMS, they will process node join request, build the IP prefix to ASN mapping table, build ASN to CN mapping table, build the partial SNs table and administrate the delay matrices. With publicly available BGP tables and updates, such as RouteViews (http://www.routeviews.org/), it is easy to obtain the IP prefix to ASN mapping table of Chinese Internet. CAIDA (http://www.caida.org/home/) can provide ASN to CN mapping table. Upon the join request of a new node, bootstraps translate the node's IP to its ASN and return the ASN to the new node. After obtaining the ASN the node can decide to join the corresponding ON-DHT overlay. In the system, every node will store the IP prefix to ASN mapping table. When the node has stored this table, it will look up the ASN by itself in the future. Bootstraps will select several SNs from each AS to build the partial SNs table. The role of this table is to facilitate the delay matrix measurement. Bootstraps will periodically generate the UDM and frequently disseminate the UDM to SNs. Since the IP prefix to ASN mapping table and ASN to CN mapping table will not change frequently, bootstraps will not frequently disseminate these two tables to SNs.

**Supernodes (SNs)**: Like the Skype, SNs are selected from ordinary nodes. SNs are more powerful and reliable nodes than ordinary nodes. SNs can take on server-like responsibilities and provide services to a set of ordinary nodes. One related SN election scheme in structured P2P network has been reported in our early paper [36]. In IDMS, SNs provide the following services. (1) Periodically contact bootstraps to retrieve the up-to-date delay matrices, the IP prefix to ASN mapping table and ASN to CN mapping table. (2) Disseminate PDM, IP prefix to ASN mapping table and ASN to CN mapping table to the ordinary nodes in their ASes. These tables and PDM will not be updated frequently. (3) Process measurement request from bootstraps, measure the path latencies between itself and SNs of other ASes, and return the measurement results to the bootstraps. (4) Accept nodal information of ordinary nodes which they serve and store top ordinary nodes. The top ordinary nodes can become SNs or take on part of job of SNs. (5) Provide specific services for ONs. (6) Generate the PDM and distribute PDM to their associated ONs.

**Ordinary nodes (ONs)**: In IDMS most nodes are ordinary nodes. They have the following duties. (1) Connect to at least two SNs in their AS enhance the reliability. (2) Retrieve PDMs, IP prefix to ASN mapping table and ASN to CN mapping table from their associated SNs. (3) Run the SN election algorithm to elect the partial powerful hosts. (4) Periodically publish their nodal information to their associated SNs.

### 4.2.4 Redundancy design

To enhance the system reliability and to avoid a single point of failure of SN, redundancy mechanism is introduced into the design of IDMS. Each ON is associated to at least two SNs. We say that an ON is *k-redundancy* if it connects to *k* SNs. A *k-redundancy* design has much greater availability and reliability. Since a SN may become a potential bottleneck and a single point of failure. When the SN simply quits or fails, all the ONs which the SN serves become temporary lost connection until they can find a new SN to connect to. If one ON connects to *k* SNs, all the associated SNs can respond to queries from this ON. If one SN fails, the others may continue to serve the ON until a new SN is selected. The probability that all associated SNs will fail before any new SNs are selected is much lower than the probability of a SN failure. In the design of IDMS, *k=2*.

In IDMS, all the ONs in the same AS form the ON-DHT to enhance the redundancy and relieve the load of SNs. For example, the resources can be simultaneously stored in the SN-DHT and the ON-DHT to realize redundancy. For one VoIP session occurred in the same AS, the ON-DHT can also process this session.

## 4.3 Delay Matrix Construction Method

In the design of IDMS, bootstraps will generate one UDM per hour. The specific procedures are listed below:

**Step 1:** The bootstraps will select one or more SNs in each AS from the partial SNs table, then simultaneously send the measurement request messages to all the selected SNs.

**Step 2:** For each destination AS, the selected SNs will measure the RTTs to one or several SNs of that AS.

**Step 3:** The selected SNs will return the measurement results to the bootstraps.

**Step 4:** Finally when the bootstraps receive all the measurement results, they will generate the UDM.

Note that the SNs can also distribute the measurement task to the ordinary nodes to relieve the load on itself and speed up the measurement process.

The PDM is generated in the SNs. The SNs will use the median of previous delay matrices to generate the PDM. For example, if one path latencies were 51ms, 55ms and 63ms separately in three previous delay matrices with the same time period, then 55ms was used in the previous delay matrix.

## 4.4 Broadcast Mechanism of Delay Matrix

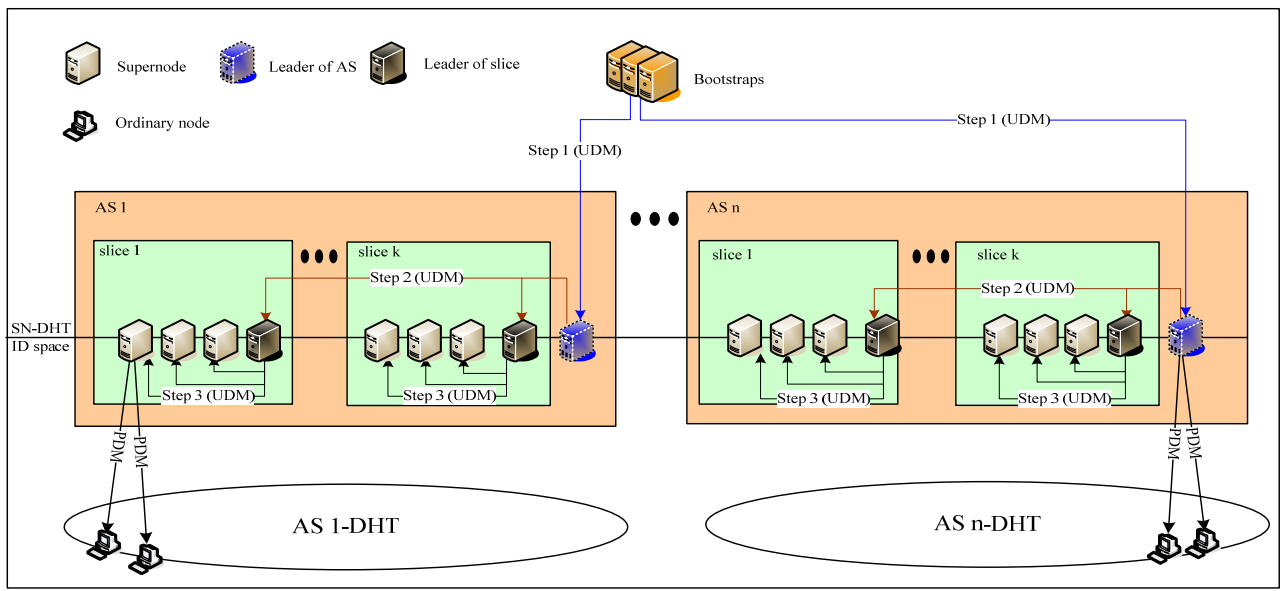

**Fig.5. Broadcast mechanism of delay matrix**

Figure 5 shows the application-level broadcast mechanism used to distribute the UDM to all the SNs in the system. The broadcast mechanism is based on two-tier hierarchical architecture. Each AS independently distributes the delay matrix. For one AS with m SNs, the first SN of the AS in the ID space of SN-DHT is selected as the AS leader. The SNs within the same AS are evenly divided into k slices which averagely contain about m/k SNs since the property of consistent hash function [24] can guarantee that node ID can be evenly distributed throughout the ID space. In each slice the first SN in ID space is selected as the slice leader. Since the SNs form a one-hop DHT and every SN has the full route table, every SNs can easily know which slices it belongs to and whether or not it is slice leader or AS leader.

The flows of broadcast mechanism of UDM are listed below:

Step 1: The bootstraps send the UDM to all the AS leaders in the systems.

Step 2: The AS leaders send the UDM to all the slice leaders in their corresponding ASes.

Step 3: The slice leaders send the UDM to all the SNs in their corresponding slices.

Finally every SN in the system obtains the UDM. The broadcast mechanism is over. Our measurement study showed that the delay matrices with the same time period are highly similar, so the bootstraps can only update the latencies with large variations rather than the whole delay matrix. Thus the traffic load can be significantly reduced.

The flow of broadcast mechanism of PDM is very simple. Each SN will distribute the PDM to every ON which it serves. Note that PDM will not be updated frequently. For example, PDM can be updated once several days or one week.

# 5. Performance Evaluation

We compare IDMS with Phoenix [23]. To our best knowledge, Phoenix is the most accurate NCS so far. For Phoenix, it uses 10-dimensional coordinates and each host has 32 reference hosts. The other parameters are the default parameters of Phoenix. We select two network scenarios of Chinese Internet. The first scenario represents the least congested network scenario and is called LCN in this paper. The corresponding delay matrix was measured between 5:00 and 6:00. The average RTT in LCN is about 44ms. The second scenario represents the most congested network scenario and is called MCN in this paper. The corresponding delay matrix was measured between 21:00 and 22:00. The average RTT in MCN is about 66ms. We select these delay matrices from day 1 to day 4. We assume the delay matrices of day 4 as the UDMs. The PDMs are computed from the delay matrices of day 1 to day 3. For Phoenix, ten independent runs are conducted on UDM and the average results are reported.

The following five aspects are considered: (1) Prediction accuracy; (2) TIV Accuracy; (3) Cost analysis; (4) Traffic load and scalability analysis; (5) Feasibility analysis.

## 5.1 Prediction Accuracy

To evaluate the accuracy of IDMS, we compared it against Phoenix which runs on UDM. Relative error (RE) defined in formula 2 is widely used to evaluate the accuracy of NCS. We also select three independent delay matrices with the same time period from day 1 to day 3 for comparison. These delay matrices are called day 1, day 2 and day 3 respectively. In the comparison UDM is set as the benchmark and there is no prediction error in UDM. The mean, median and 90th percentile relative error (NPRE) [22-23] are used to evaluate prediction accuracy. Smaller NPRE value means higher overall prediction accuracy.

Figure 6 and figure 7 show the relative errors of all links in LCN and MCN. The detailed results are showed in table 2. Compared with Phoenix, IDMS using PDM reduces the NPRE by between 76.56% (LCN) and 53% (MCN), the mean by between 77.65% (LCN) and 72.15% (MCN) and the median between 92.31% (LCN) and 92.31% (MCN). In LCN and MCN, IDMS using PDM achieves significantly higher prediction accuracy than Phoenix. The results of three independent delay matrices are slightly worse than that of IDMS using PDM, so in IDMS the median value is used in PDM which is computed from several previous delay matrices measured in the same time. Since the median relative error of IDMS using PDM is only 1% and 2% in LCN and MCN respectively, it means that for at least 50% of links the prediction errors can be ignored.

**Table 2** Relative errors

|  |  | Phoenix | day 1 | day 2 | day 3 | IDMS using PDM | IDMS using UDM |
|---|---|---|---|---|---|---|---|
| All links(LCN) | NPRE | 0.64 | 0.19 | 0.18 | 0.18 | 0.15 | 0 |
|  | Mean | 0.85 | 0.27 | 0.31 | 0.25 | 0.19 | 0 |
|  | Median | 0.13 | 0.01 | 0.01 | 0.01 | 0.01 | 0 |
| All links(MCN) | NPRE | 1.00 | 0.57 | 0.58 | 0.58 | 0.47 | 0 |
|  | Mean | 0.79 | 0.40 | 0.47 | 0.51 | 0.22 | 0 |
|  | Median | 0.26 | 0.04 | 0.03 | 0.03 | 0.02 | 0 |
| Short links(LCN) | NPRE | 0.85 | 0.15 | 0.14 | 0.14 | 0.11 | 0 |
| | Mean | 1.18 | 0.36 | 0.41 | 0.32 | 0.24 | 0 |
| | Median | 0.14 | 0.01 | 0.01 | 0.01 | 0.01 | 0 |
| Short links(MCN) | NPRE | 1.69 | 0.35 | 0.33 | 0.34 | 0.24 | 0 |
| | Mean | 1.09 | 0.45 | 0.56 | 0.60 | 0.22 | 0 |
| | Median | 0.25 | 0.03 | 0.02 | 0.02 | 0.02 | 0 |

We further evaluate the prediction accuracy of short links (less than 50ms). For many Internet distributed systems, short links are very important to select close-by hosts [6] and relay paths [4]. Figure 8 and figure 9 show the relative errors of short links in LCN and MCN. The detailed results are showed in table 2. The results of short links are similar to that of all links. Specifically, compared with Phoenix, IDMS using PDM reduces the NPRE by between 87.06% (LCN) and 85.80% (MCN), the mean by between 79.66% (LCN) and 79.82% (MCN) and the median between 92.86% (LCN) and 92.00% (MCN). For short links, IDMS using PDM also achieves much higher prediction accuracy than Phoenix. The results of three independent delay matrices are slightly worse than that of IDMS using PDM. One interesting result is that in LCN and MCN the NPREs of short links of IDMS using PDM are smaller than that of all links of IDMS using PDM. It means that for short links IDMS using PDM can provide more accurate prediction. However, the NPREs of short links of Phoenix are larger than that of all links of Phoenix in LCN and MCN. In short, IDMS using PDM achieves significantly lower relative error than Phoenix, especially for short links.

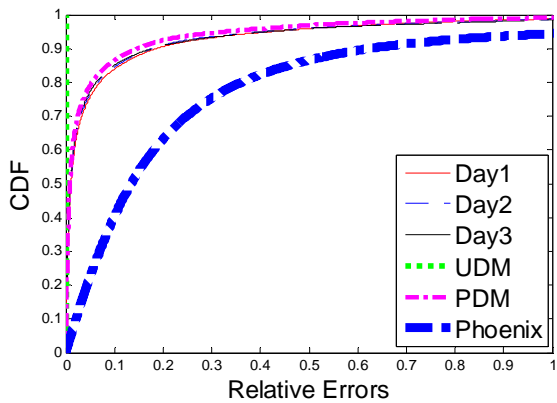
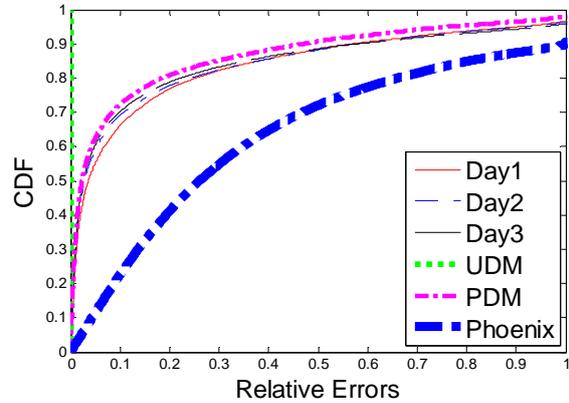

**Fig.6. CDF of relative errors of all links in LCN**  **Fig.7. CDF of relative errors of all links in MCN**

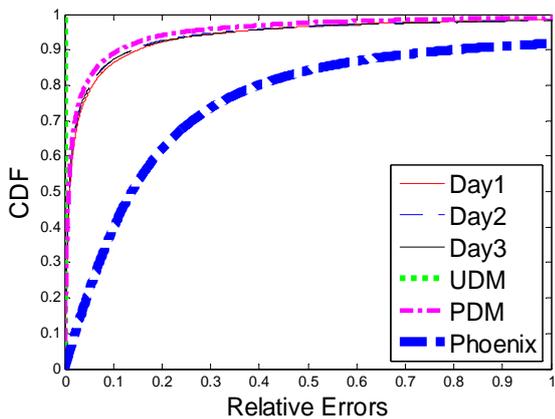
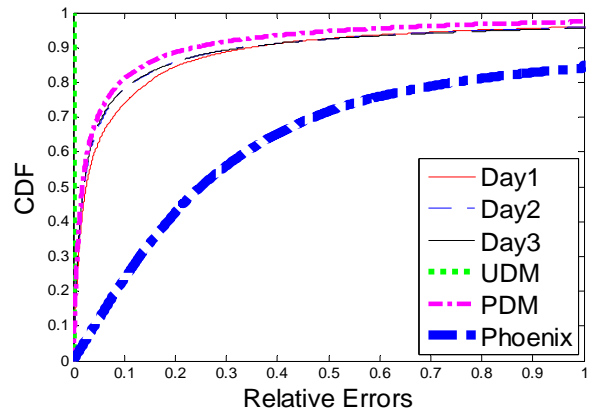

**Fig.8. CDF of relative errors of short links in LCN**  **Fig.9. CDF of relative errors of short links in MCN**

### 5.2 TIV Accuracy

Many studies have reported that TIV is a common phenomenon on the Internet. Though TIV is harmful for most NCSes, TIVs expose opportunity to find a fast alternate path for numerous delay sensitive applications to improve their performances.

Therefore, IDMS and Phoenix should reserve as many TIVs as possible. Since in IDMS it directly uses the delay matrix and does not further process the delay matrix, IDMS using UDM can perfectly retain the feature of TIV. In the comparison UDM is set as the benchmark. We compared IDMS using PDM against Phoenix which runs on UDM.

TIV victory (formula 3) and TIV failure (formula 4) are proposed to evaluate TIV accuracy.

$$TIV\_V = \frac{\text{the number of TIVs (in PDM or Phoenix) which occure in UDM}}{\text{the total number of TIVs in UDM}} \quad (3)$$

$$TIV\_F = \frac{\text{the number of TIVs (in PDM or Phoenix) which do not occure in UDM}}{\text{the total number of TIVs in UDM}} \quad (4)$$

Smaller $TIV\_F$ and bigger $TIV\_V$ indicate higher TIV accuracy. In the comparison, we define two types of TIVs. Original TIV (OTIV) is defined as $L(A,B) > L(A,C) + L(C,B)$. Practical TIV (PTIV) is defined as $L(A,B) > L(A,C) + L(C,B) + 40\text{ms}$. 40ms is added to the relay paths ($L(A,C), L(C,B)$) for considering relay delay [4]. It means that the practical relay paths must save at least 40ms than the direct paths. Figure 10 shows the OTIV accuracy in LCN and MCN. Specifically, compared with Phoenix, IDMS using PDM enhances the $TIV\_V$ by between 51.91% (LCN) and 70.56% (MCN) and reduces the $TIV\_F$ by between 57.26% (LCN) and 53.71% (MCN). In LCN and MCN, IDMS using PDM achieves significantly higher prediction accuracy than Phoenix. However, the $TIV\_V$ and $TIV\_F$ of Phoenix in MCN is comparability which may be hard to distinguish real TIVs and false TIVs.

Figure 11 shows the PTIV accuracy in LCN and MCN. The results of PTIV accuracy are similar to that of OTIV accuracy. Specifically, compared with Phoenix, IDMS using PDM enhances the $TIV\_V$ by between 211.69% (LCN) and 208.56% (MCN) and reduces the $TIV\_F$ by between -9.27% (LCN) and 11.29% (MCN). In LCN and MCN, the $TIV\_V$ and $TIV\_F$ of Phoenix is comparability which may be hard to distinguish real TIV and false TIV. It means that Phoenix can not perfectly retain the TIV property. In short, IDMS using PDM achieves significantly higher TIV accuracy than Phoenix, especially for PTIV.

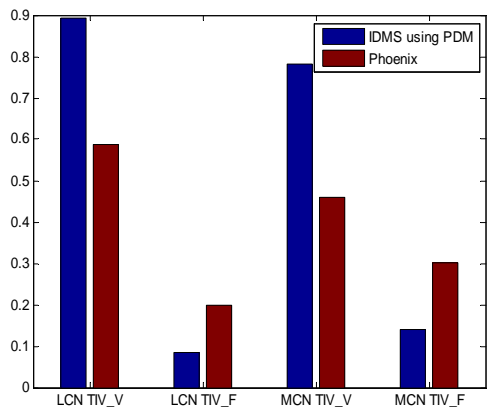
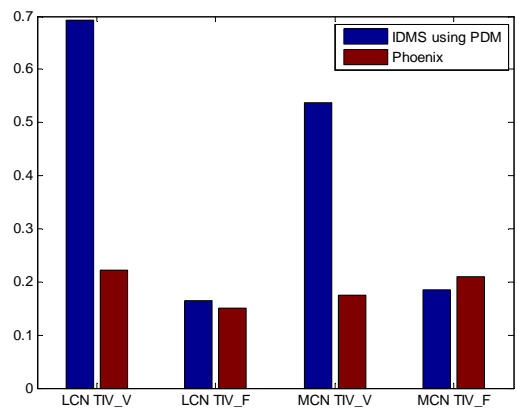

**Fig.10. OTIV Accuracy**  **Fig.11. PTIV Accuracy**

### 5.3 Cost analysis

The cost in IDMS is mainly composed of message overhead and storage overhead. We will separately analyze. In the analysis we assume there are $N$ nodes in the system. The ratio of SNs in IDMS is $p$. Each SN serves about $1/p$ ONs. The measurement time interval is one hour in IDMS. The number of ASes is $L$. The size of delay matrix in IDMS is $LxL$. For fair comparison, each host in Phoenix will update its coordinate once in one hour. For UDM, Phoenix requires at least 6 rounds to reach steady state. In this subsection, we only compare the cost of IDMS using UDM with that of Phoenix since PDM will not be frequently updated.

**Message overhead:** The message overhead in IDMS is composed of messages in the delay matrix construction and messages in the distribution of delay matrix. We use $m$ bytes to describe an event and $b$ bytes to describe the overhead per message. The latency of one path is described as one event. In delay matrix construction (section 4.3), we assume that the bootstraps will select one SNs in each AS and each selected SN will measure one SN in any other AS. The IDMS will generate $2L(L+1)$ messages in one hour in delay matrix construction. The size of these messages is $3L(m+b)+2L(m+b)(L-1)+ L(b+m(L-1))$. In the broadcast of delay matrix (section 4.4), IDMS will generate $2Np$ messages. The size of these messages is $Np(m+b)+Npzq$, where $z$ is the file size of UDM and $q$ is the ratio of data update required in UDM.

For Phoenix, each host has 32 reference hosts. In one round, Phoenix will generate *64N* messages. To reach steady state, Phoenix will generate *384N* messages. The size of these messages is *384N (m+b)*.

In the analysis *p=0.01, z=2L$^2$* (2 bytes to store one latency value), *q=0.1, m=20, b=40*, setting *L* as 50, 100, 200, 555 separately where 555 is the number of ASes in Chinese Internet till Jun, 2011 (http://ipwhois.cnnic.cn/ipstats/index.php?obj=asn). Figure 12 shows the number of messages. Figure 13 shows the size of messages. We see that the message overhead of IDMS using UDM is much smaller than that of Phoenix.

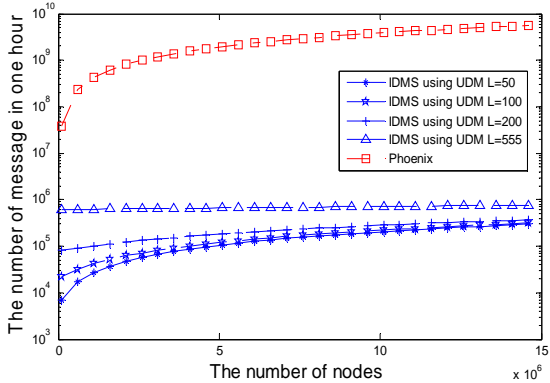 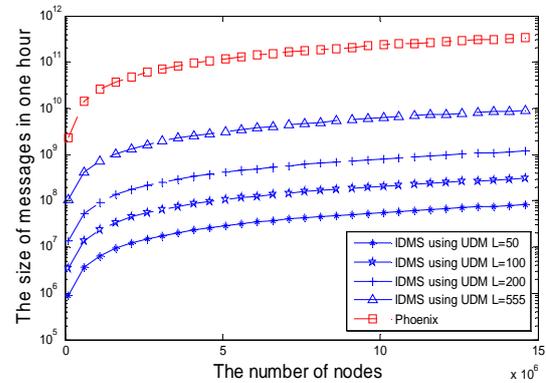

**Fig.12. The number of messages**    **Fig.13. The size of messages**

**Storage overhead:** In IDMS each SN needs to store the UDM. It will take about 500 KB to store one 555x555 delay matrix. For each SN there will be one UDM in the memory. For each ON there will be one PDM in the memory. Considering most computers will have 500G hard drives and 2G memory, the memory required will not be a problem for any SN or ON in IDMS since it only takes 0.025% memory. If the measurement interval is one hour, IDMS will generate 24 delay matrices in one day. The storage required will not be a problem for any SN or ON since the storage required is about 12MB.

### 5.4 Traffic Load and Scalability Analysis

On the bootstraps and SNs, they will process several hundreds messages in the process of the delay matrix construction and in the distribution process of delay matrix. The traffic load on bootstraps and SNs should be very small.

For IDMS even the bootstraps are failure for a long time, IDMS can also achieve very high performance since IDMS using PDM is also very accurate. Since the size of delay matrix is only related to the number of ASes in Chinese Internet and the analysis shows the cost is very small and the traffic load on any node may not be a problem, IDMS can scale to a very large network. There is not scalability problem in IDMS for Chinese Internet.

### 5.5 Feasibility Analysis of IDMS

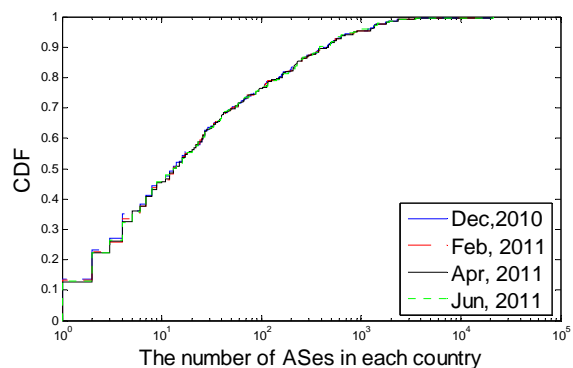

**Fig.14. The distribution of ASes in each country**

We have analyzed that IDMS is an efficient approach for Chinese Internet. Does IDMS is feasible in other countries? This problem is also very important. Since IDMS is related with the number of ASes in each country, we analyze the distribution of ASes in each country. Figure 14 plots the distribution of ASes in each country from Dec, 2010 to Jun, 2011

(http://ipwhois.cnnic.cn/ipstats/index.php?obj=asn). We see that over 90% of countries the number of ASes is smaller than that of China. Our analysis shows that IDMS is more efficient in the countries with small number of ASes. So IDMS is feasible in most countries.

## 6. Conclusion and Future Work

In this paper, we have presented a new approach called IDMS based on delay matrix for predicting network distances between arbitrary Internet hosts. Different from most other NCSes which treat Internet as a black box, IDMS exploits the properties of delay space of Chinese Internet. Since the hosts in the same AS are close to each other and TIVs seldom occur in the same AS, it is not need to assign each host an independent NC. In IDMS hosts in the same AS are assigned the same locality ID. Therefore, the locations of hosts can be identified by locality ID. For the network distances among ASes, IDMS periodically generates the delay matrix of ASes and directly use the delay matrix. Therefore, it can significantly improve the prediction accuracy and retain TIV property. Since the delay matrices with the same time period are highly similar in every day, the previous delay matrices can be used to estimate the future delay matrices with the same time period. Exploiting this property IDMS introduces two types of delay matrices. One is the UDM which is the most accurate delay matrix. The other is the PDM which also can provide high prediction accuracy. Extensive simulations on our collected Chinese Internet data sets show that IDMS is an accurate, efficient, scalable and practical method for Chinese Internet. The performance of IDMS is better than other existing approaches.

The IDMS are very helpful for close-by nodes selection and relay node selection. For example, close nodes in the same AS can be identified by locality ID. Close nodes in the other ASes can be directly looked up from the UDM or PDM. Relay node selection in other ASes can be directly looked up from the UDM or PDM. There are some open problems to be addressed. For example, we do not consider the security and authentication related issues. When there are some malicious nodes that misreport the path latencies in the IDMS, the performance of IDMS may be degradation. How to design IDMS for unstructured P2P networks? How to design IDMS which can cover many countries? How many reference points are suitable for a large AS in the delay matrix? For a country with large number of ASes, the size of delay matrix of IDMS will be very large. The cost of IDMS is also very large. How to solve this problem? These problems will be our future work. We also plan to implement IDMS in our own P2PSIP-VoIP system for further evaluation.

## Acknowledgements

We thank anonymous reviewers for their constructive suggestions. Thanks to Yang Chen for providing the source code of Phoenix. This work is supported by the International Scientific and Technological Cooperation Projects under No. 2010DFA12780, the Key National Science & Technology Specific Projects under No. 2010ZX03005-003 and the Fundamental Research Funds for the Central Universities under No. 2009RC0121.